\documentclass{llncs}
\usepackage{llncsdoc}
\usepackage{graphicx}
\begin{document}
\newcounter{save}\setcounter{save}{\value{section}}
{\def\addtocontents#1#2{}%
\def\addcontentsline#1#2#3{}%
\def\markboth#1#2{}%
\title{Phase Transition of a Skeleton Model for Surfaces}
\author{Hiroshi Koibuchi\inst{1}}

\institute{Ibaraki National College of Technology, Nakane 866, Hitachinaka, Ibaraki 312-8508, Japan}

\maketitle
\begin{abstract}
A spherical model of skeleton with junctions is investigated by Monte Carlo simulations. The model is governed by one-dimensional bending energy. The results indicate that the model undergoes a first-order transition separating the smooth phase from the crumpled phase. The existence of phase transition indicates that junctions play a non-trivial role in the transition.   
\end{abstract}
\section{Introduction}
Surface model of Helfrich, Polyakov and Kleinert (HPK) \cite{HELFRICH-1973,POLYAKOV-NPB1986,Kleinert-PLB1986} has long been investigated for a model of biological membranes \cite{NELSON-SMMS2004,Gompper-Schick-PTC-1994,Bowick-PREP2001,Peliti-Leibler-PRL1985,DavidGuitter-EPL1988,PKN-PRL1988}. A curvature energy called the bending energy is assumed in the Hamiltonian, which is discretized on triangulated surfaces in numerical studies \cite{KANTOR-NELSON-PRA1987}. It was reported that the model undergoes a first-order transition between the smooth phase and a crumpled (or wrinkled) phase \cite{KD-PRE2002,KOIB-PRE-2005,KOIB-NPB-2005}. Experimentally, a similar transition was recently observed in an artificial membrane \cite{CVNA-PRL-2006}.

The curvature energy in HPK model is a two-dimensional one, and it reflects a homogeneous structure of membranes as a two-dimensional surface. In fact, the conventional picture of biological membranes is connected to such homogeneity, and moreover many artificial membranes are known as homogeneous.

 However, membrane skeleton, called the cytoskeleton in biological membranes,  has been considered to play a crucial role in maintaining shape and motion. Recent experimental studies reveal that free diffusion of lipids suffers from the existence of cytoskeletons: the so-called hop-diffusion was actually observed \cite{Kusumi-BioJ-2004}. The artificial membrane, in the above-mentioned experimental study, is considered not to be a homogeneous surface, because it is partly polymerized \cite{CVNA-PRL-2006}.   

Therefore, it is interesting to study whether the phase transition occurs in a skeleton model, which is obviously different from the conventional surface model such as HPK model. The skeleton model is significantly simplified as a model for surfaces; it contains only skeletons and the junctions. Consequently, we consider that the simplified model enable us to focus on whether skeletons play a crucial role on the phase transition in biological and artificial membranes.

In this paper, we report numerical evidence that a phase transition occurs in a skeleton model, which is governed by one-dimensional bending energy. This result is remarkable, because no phase transition can be observed in one-dimensional object described by Hamiltonian of local interactions.

\section{Model and Monte Carlo technique}

By dividing every edge of the icosahedron into some pieces of the uniform length, we have a triangulated surface. The compartment structure is built on the surface by eliminating the vertices inside the compartment. The boundary of the compartment forms the skeletons with junctions. We denote by $(N,N_S,N_J,L)$ the total number of vertices (including those in the junctions and their neighboring vertices), the total number of vertices in the skeletons (excluding both vertices in the junctions and their neighboring vertices), the total number of junctions, and the number of bonds between two neighboring junctions in a compartment, respectively. Note that $L-1$ is the number of vertices in the skeleton between the junctions. 

A starting configuration of Monte Carlo (MC) simulation is shown in Fig. \ref{fig-1}(a), \ref{fig-1}(b), where $(N,N_S,N_J,L)\!=\!(2562,1440,162,4)$ and $(N,N_S,N_J,L)\!=\!(19362,10890,1212,4)$. The dots are clearly seen in (a) and unclear in (b) because the size of (b) is relatively large. The surface of the sphere is shown in the snapshot for clarity. The hexagon and the pentagon (which are clearly seen in (a)) correspond to the junctions. Total number of the pentagon is 12, and the remaining junctions are hexagonal. 
%++++++++++++++++++++++++++++++++++
\begin{figure}[htb]
%\vspace{1cm}
%WinTpicVersion3.08
\centering
%WinTpicVersion3.08
\unitlength 0.1in
\begin{picture}( 0,0)(  10,10)
\put(16,8.5){\makebox(0,0){(a) $(2562,1440,162,4)$ }}%
\put(38,8.5){\makebox(0,0){(b) $(19362,10890,1212,4)$ }}%
\end{picture}%
\vspace{0.5cm}
\includegraphics[width=10cm]{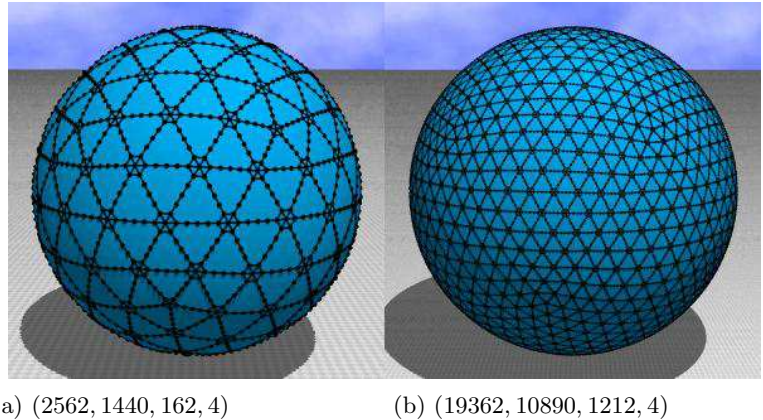}
\caption{Starting configuration of surfaces of (a) $(N,N_S,N_J,L)\!=\!(2562,1440,162,4)$ and (b) $(N,N_S,N_J,L)\!=\!(19362,10890,1212,4)$. Thick lines denote the skeletons and the junctions, and small dots the vertices. The dots are clearly seen in (a) and unclear in (b) because the surface size is relatively large. The surface of the sphere is shown for clarity in both (a) and (b). } 
\label{fig-1}
\end{figure}
%++++++++++++++++++++++++++++++++++

Henceforth, we use terminology {\it junction} for the hexagonal (or pentagonal) objects or for the central vertices of these objects, {\it skeleton} for linear object between junctions, and {\it surface} for junction $+$ skeleton.

The Gaussian bond potential $S_1$, the one-dimensional bending energy $S_2^{(1)}$ for skeletons, and the two-dimensional bending energy $S_2^{(2)}$ for junctions, are given by
\begin{equation}
\label{Disc-Eneg} 
S_1=\sum_{(ij)} \left(X_i-X_j\right)^2,\quad S_2^{(1)}=\sum_{[ij]} (1-{\bf t}_i \cdot {\bf t}_j),\quad 
S_2^{(2)}=\sum_{\langle ij\rangle} (1-{\bf n}_i \cdot {\bf n}_j),
\end{equation} 
where $\sum_{(ij)}$ in $S_1$ is the sum over bonds $(ij)$ connecting all the vertices $i$ and $j$ (including those at the junctions and their neighboring vertices), $\sum_{[ij]}$ in $S_2^{(1)}$ is the sum over nearest neighbor bonds $i$ and $j$ on the skeletons between junctions, $\sum_{\langle ij\rangle}$ in $S_2^{(2)}$ is the sum over bonds $\langle ij\rangle$ connecting the central point of the junction $i$ and the nearest neighbor vertex $j$. The symbol ${\bf t}_i$ in $S_2^{(1)}$ is the unit tangential vector of the bond $i$. The symbol ${\bf n}_i$ in $S_2^{(2)}$ is the unit normal vector of the triangle $i$.

 $S_2^{(2)}$ is defined only at the junctions.  Note also that the bond $[ij]$ in $S_2^{(1)}$ includes the bonds connected to the central point of the junction. $S_1$ includes bonds that connect the vertices nearest neighbor to the junctions. The reason of this is for the sake of the in-plane elasticity at the junctions. When $S_1$ excludes such bonds, skeletons at the junctions freely move into the in-plane directions. The potential $S_1$ describes a one-dimensional interaction between vertices in the skeletons and a two-dimensional interaction only at the junctions.

The partition function of the model is defined by
\begin{equation} 
\label{Part-Func}
 Z = \int^\prime \prod _{i=1}^{N} d X_i \exp(-S),\quad
 S=S_1 + b_1 S_2^{(1)} + b_2 S_2^{(2)}, 
\end{equation} 
where $b_1$, $b_2$ are the one-dimensional bending rigidity and the two-dimensional bending rigidity. In this paper, $b_2$ is fixed to $b_2\!=\!5$, and $b_1$ is varied in the MC simulations. The center of the surface is fixed in the MC simulations, and this is denoted by $\prime$ in $\int^\prime \prod _{i=1}^{N} d X_i$. 

The canonical Metropolice MC technique is used to obtain the mean value of physical quantities. The vertices $X$ of the skeleton are sequentially shifted so that $X^\prime \!=\! X\!+\!\delta X$, where $\delta X$ is randomly chosen in a small sphere. The new position $X^\prime$ is accepted with the probability ${\rm Min}[1,\exp(-\Delta S)]$, where $\Delta S\!=\! S({\rm new})\!-\!S({\rm old})$. The vertices of junctions (hexagons and pentagons) are shifted in two steps. Firstly 7 (or 6) vertices at one junction are shifted simultaneously at random, and the second step is a random translation and a random rotation of the hexagon (or the pentagon) as a rigid object. Those shifts are controlled by small numbers given at the beginning of the simulations to maintain about $50\%$ acceptance rates.

Four different sizes are assumed for surfaces in the MC simulations. The 
surface size $(N,N_S,N_J,L)$ are $(19362,10890,1212,4)$, $(10242,5760,642,4)$, \\$(5762,3240,362,4)$, and $(2562,1440,162,4)$. In the simulations, two parameters are fixed: the total number of vertices $L$ in the skeleton between junctions and the two-dimensional bending rigidity $b_2$ are fixed to
\begin{equation} 
\label{parameters}
L=5,\quad b_2=5.
\end{equation}

\section{Results}
%++++++++++++++++++++++++++++++++++
\begin{figure}[htb]
\centering
%\vspace{1cm}
%WinTpicVersion3.08
\unitlength 0.1in
\begin{picture}( 0,0)(  10,10)
\put(16,43.5){\makebox(0,0){(a) $b_1\!=\!2.9$, crumped }}%
\put(34,43.5){\makebox(0,0){(b) $b_1\!=\!2.93$, smooth }}%
\put(18,8.5){\makebox(0,0){(c) the surface section of (a) }}%
\put(37,8.5){\makebox(0,0){(d) the surface section of (b) }}%
\end{picture}%
\vspace{0.5cm}
%\centering
\includegraphics[width=8cm]{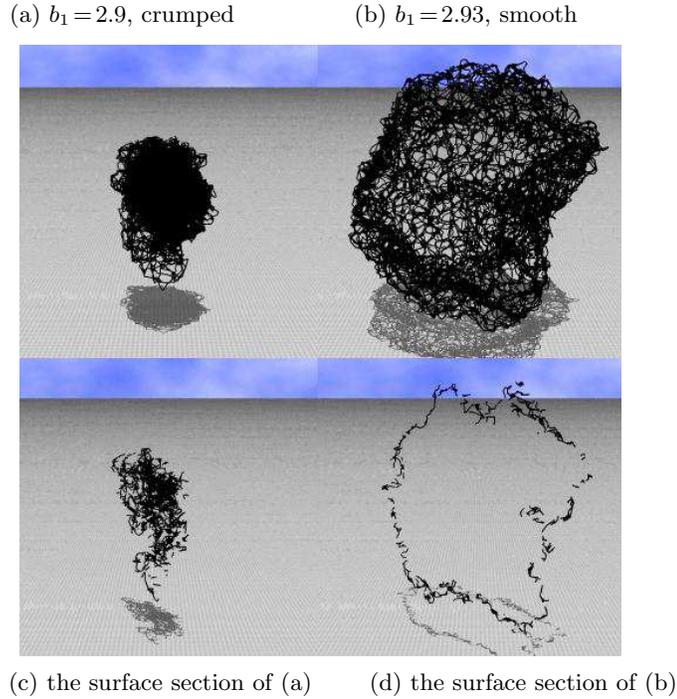}
\caption{Snapshots of surfaces of size $(N,N_S,N_J,L)\!=\!(19362,10890,1212,4)$obtained at (a) $b_1\!=\!2.9$ (crumpled) and  (b) $b_1\!=\!2.93$ (smooth). (c) The surface section of (a), and (d) the surface section of (b). } 
\label{fig-2}
\end{figure}
%++++++++++++++++++++++++++++++++++
First, we show snapshots of surface obtained at $b_1\!=\!2.9$ (crumpled phase) and $b_1\!=\!2.93$ (smooth phase) in Figs.\ref{fig-2}(a) and \ref{fig-2}(b), respectively. The size of surface is $(N,N_S,N_J,L)\!=\!(19362,10890,1212,4)$. The surface sections are shown in  Figs.\ref{fig-2}(c),\ref{fig-2}(d). We clearly find that the surface is crumpled (smooth) at $b_1\!=\!2.9$ ($b_1\!=\!2.93$).  

%++++++++++++++++++++++++++++++++++
\begin{figure}[htb]
\centering
\includegraphics[width=10cm]{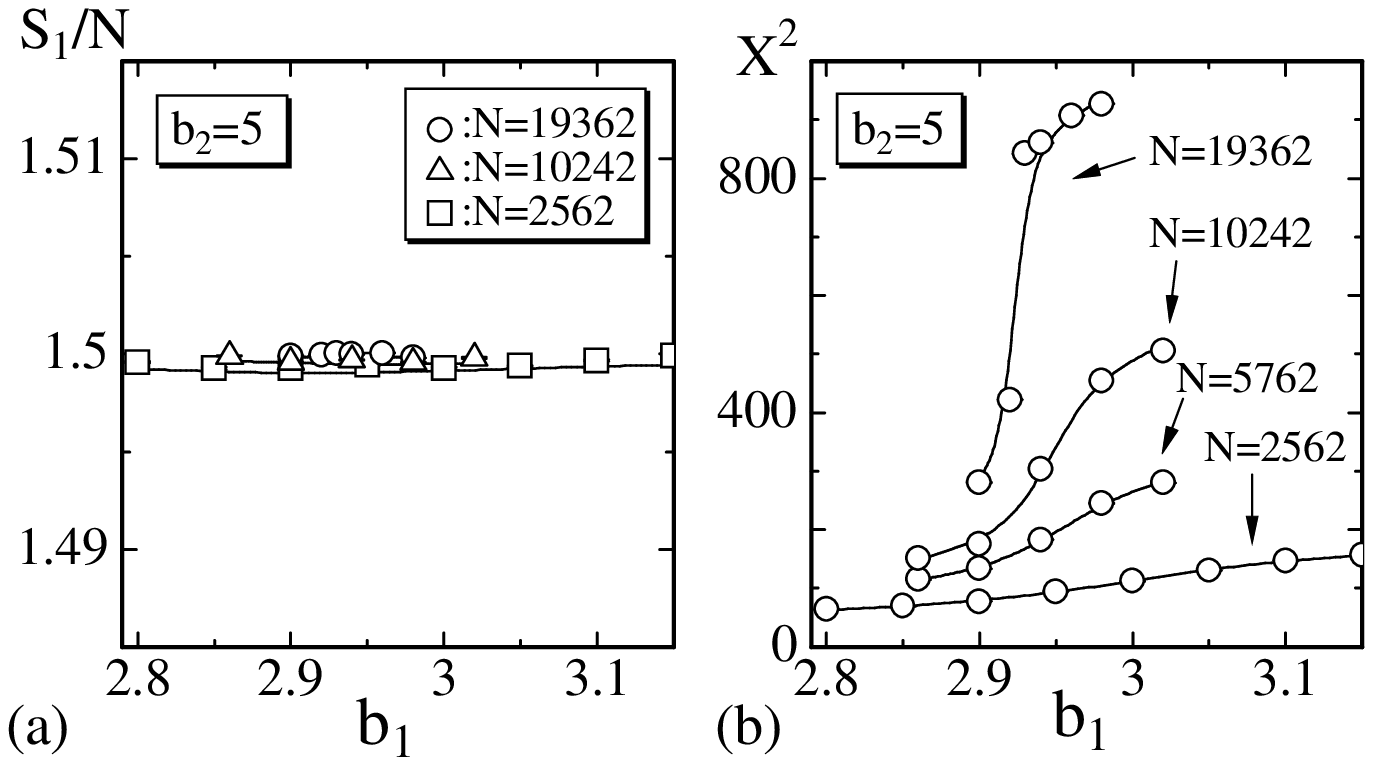}
\caption{(a) $S_1/N$ against $b_1$, and (b) $X^2$ against $b_1$. $b_2$ is fixed to $b_2\!=\!5$. The curves are drawn by multi-histogram reweighting technique.} 
\label{fig-3}
\end{figure}
%++++++++++++++++++++++++++++++++++
Figure \ref{fig-3}(a) is plots of $S_1/N$ against $b_1$ obtained on the surfaces of size  
$(19362,10890,1212,4)$,  $(5762,3240,362,4)$, and $(2562,1440,162,4)$. Because of the scale invariant property of the partition function $Z$ in Eq.(\ref{Part-Func}), $S_1/N$ is predicted to be $S_1/N\!=\!3N/2(N-1)\!\simeq\!1.5$. We find from the figure that the predicted relation for $S_1/N$ is satisfied. The line connecting the data was obtained by multi-histogram reweighting technique.

Figure \ref{fig-3}(b) is the mean square size $X^2$ defined by
\begin{equation}
\label{X2}
X^2={1\over N} \sum_i \left(X_i-\bar X\right)^2, \quad \bar X={1\over N} \sum_i X_i,
\end{equation}
where $\bar X$ is the center of the surface. The size  $X^2$ of surfaces increases with increasing $b_1$ and varies rapidly at intermediate $b_1$ on the surface of \\$(19362,10890,1212,4)$. This indicates the existence of the crumpling transition.

%++++++++++++++++++++++++++++++++++
\begin{figure}[hbt]
\centering
\includegraphics[width=10cm]{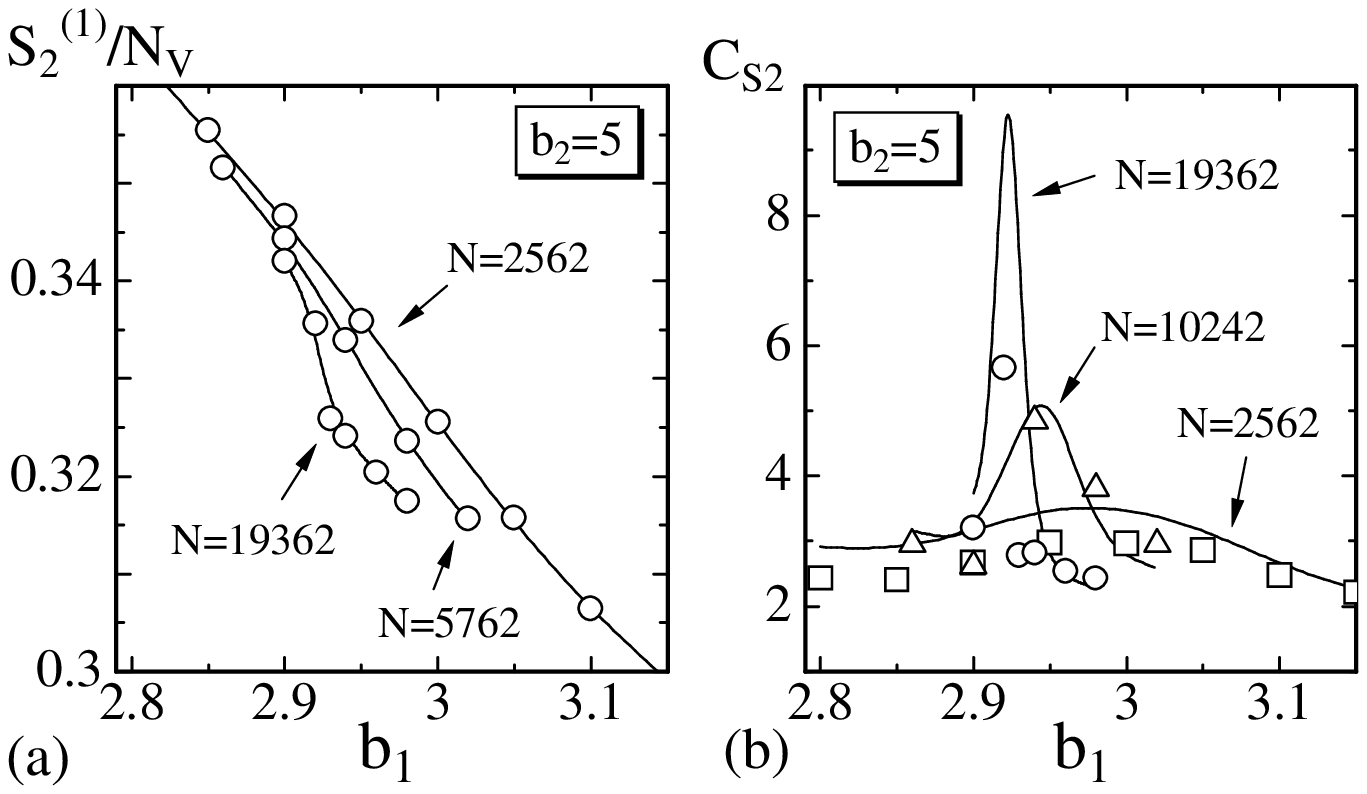}
\caption{(a) $S_2^{(1)}/N_V$ against $b_1$, and (b) $C_{S_2^{(1)}}$ against $b_1$. $N_V\!=\!N\!-\!N_J$ is the total number of vetices where $S_2^{(1)}$ is defined. $b_2$ is fixed to $b_2\!=\!5$.  The curves are drawn by multi-histogram reweighting technique.} 
\label{fig-4}
\end{figure}
%++++++++++++++++++++++++++++++++++
The bending energy $S_2^{(1)}/N_V$ against $b_1$ is plotted in Fig. \ref{fig-4}(a), where $N_V\!=\!N\!-\!N_J$ is the total number of vetices where $S_2^{(1)}$ is defined. We find again a rapid varying of $S_2^{(1)}/N_V$ against $b_1$ as the surface size increases. 

Figure \ref{fig-4}(b) shows the specific heat $C_{S_2^{(1)}}$ defined by     
\begin{equation}
C_{S_2^{(1)}} \!=\! {b^2\over N_V} \langle \; \left( S_2^{(1)} \!-\! \langle S_2^{(1)} \rangle\right)^2\rangle. 
\end{equation}
The curves drawn by multi-histogram reweighting technique indicate an anomalous behavior of $C_{S_2^{(1)}}$. This can be considered as a sign of the phase transition. We expect that the transition is of first-order because of the sharp peak in $C_{S_2^{(1)}}$ at the surface of $(N,N_S,N_J,L)\!=\!(19362,10890,1212,4)$. The junctions play a nontrivial role in the phase transition, because we know that one-dimensional linear skeleton such as an elastic circle has no phase transition.  The finite-size scaling analysis can be done with more extensive MC simulations including larger surfaces rather than those in this paper, and the first-order transition will be confirmed. 

\section{Summary and Conclusion}
We have investigated the phase structure of a skeleton model for membranes by using Monte Carlo simulation technique. Linear skeletons are joined with junctions and form a surface, which covers a sphere. Hamiltonian is a linear combination of the Gaussian bond potential, one-dimensional bending energy $S_2^{(1)}$, and the two-dimensional bending energy $S_2^{(2)}$, which is defined only at the junctions. Four different sized surfaces were used in the simulation. The length $L$ of skeleton between junctions was fixed to $L\!=\!5$ in those surfaces, and the two-dimensional bending rigidity $b_2$ was also fixed to $b_2\!=\!5$ .   

We found that there are two distinct phases in the model; the smooth phase and the crumpled phase, which are characterized by small $S_2^{(1)}$ and large $S_2^{(1)}$, respectively, and also by small $X^2$ and large $X^2$. Moreover, these two phases are separated by a phase transition, which was predicted to be of first-order on account of the anomalous behavior of the specific heat for $S_2^{(1)}$. 

Although we have obtained a sign of first-order transition in the skeleton model, more extensive MC simulations are necessary to confirm the result. The junctions are expected to play a non-trivial role in the phase transition, because we know that one-dimensional linear skeleton such as an elastic circle has no phase transition. Then, it is interesting to study the dependence of junction elasticity on the phase transition. A rigid junction model is also interesting.  Rigid junction has an infinite in-plane and bending resistance against force, while the junction in this paper has finite elastic resistance. Phase structure of skeleton models can be clarified by further numerical studies. 

%\begin{acknowledgments}
This work is supported in part by a Grant-in-Aid for Scientific Research, No. 15560160.  
%\end{acknowledgments}

%----------------------------------------------------------
%\vspace*{3mm}
%\noindent
%{\bf Acknowledgment}\\
%\vspace*{2mm}
%\par

%\vfill\eject
%\vspace*{5mm}
%\noindent
%{\bf References}


\begin{thebibliography}{999}
%-----------------------------------------------
\bibitem{HELFRICH-1973}
 W. Helfrich, Z. Naturforsch, \textbf{28c} (1973) 693.

\bibitem{POLYAKOV-NPB1986}
 A.M. Polyakov, Nucl. Phys. B \textbf{268} (1986) 406.

\bibitem{Kleinert-PLB1986}
 H. Kleinert, Phys. Lett. B \textbf{174} (1986) 335.

%-----------------------------------------------
\bibitem{NELSON-SMMS2004}
D. Nelson, in \textit{Statistical Mechanics of Membranes and Surfaces, Second Edition}, edited by  D. Nelson, T.Piran, and S.Weinberg, (World Scientific, 2004), p.1. 

\bibitem{Gompper-Schick-PTC-1994}
G. Gompper and M. Schick, \textit{Self-assembling amphiphilic systems}, In
\textit{Phase Transitions and Critical Phenomena 16}, C. Domb and J.L. Lebowitz, Eds. (Academic Press, 1994) p.1.

\bibitem{Bowick-PREP2001}
 M. Bowick and A. Travesset, Phys. Rep. \textbf{344} (2001) 255.

%-----------------------------------------------
\bibitem{Peliti-Leibler-PRL1985}
 L. Peliti and S. Leibler, Phys. Rev. Lett. \textbf{54} (15)  (1985) 1690.

\bibitem{DavidGuitter-EPL1988}
 F. David and E. Guitter, Europhys. Lett,  \textbf{5} (8)  (1988) 709.

\bibitem{PKN-PRL1988}
M. Paczuski, M. Kardar, and D. R. Nelson, Phys. Rev. Lett. \textbf{60}, (1988)  2638.

%-----------------------------------------------
\bibitem{KANTOR-NELSON-PRA1987}
 Y. Kantor and  D.R. Nelson, Phys. Rev. A \textbf{36},  (1987) 4020.

%-----------------------------------------------
\bibitem{KD-PRE2002}
J-P. Kownacki and H. T. Diep, Phys. Rev. E \textbf{66},  (2002)  066105.

\bibitem{KOIB-PRE-2005}
 H. Koibuchi and T. Kuwahata, Phys. Rev. E, \textbf{72}, (2005) 026124. 

\bibitem{KOIB-NPB-2005}
 I. Endo and H. Koibuchi, Nucl. Phys. B \textbf{732} [FS], (2006) 732. 

%-----------------------------------------------
\bibitem{CVNA-PRL-2006}
Sahraoui Chaieb, Vinay K. Natrajan, and Ahmed Abd El-rahman, 
Phys. Rev. Lett. 96, (2006)  078101.

%-----------------------------------------------
\bibitem{Kusumi-BioJ-2004}
 K. Murase, T. Fujiwara, Y. Umehara, K. Suzuki, R. Iino, H. Yamashita, M. Saito, H. Murakoshi, K. Ritohie, and A. Kusumi, Biol. J. \textbf{86} (2004) 4075.

%---------------------------


\end{thebibliography}
\end{document}